\documentclass[amsfonts,prl,showpacs]{revtex4}
\begin{document}
\title{On the possible existence of crystallites in glass-forming liquids}
\author{Jeppe C. Dyre}
\address{DNRF Centre ``Glass and Time,'' IMFUFA, Department of Sciences, Roskilde University, Postbox 260, DK-4000 Roskilde, Denmark}
\date{\today}

\newcommand{\br}{{\bold r}}

\begin{abstract}
We speculate that glass-forming liquids may contain fairly large and well-defined crystallites. This is based on arguing that the slowly relaxing ``frozen-in'' stresses characterizing ultraviscous liquids increase the barrier for nucleation, thus allowing for larger unstable crystallites than otherwise possible. The frozen-in stresses also deform the crystallites, making their observation difficult; specifically it is argued that a situation where $1/N$ of the molecules form $N\times N\times N$ crystallites would be hard to detect by standard X-ray or neutron scattering experiments.
\end{abstract}


\maketitle

The glass transition takes place when a liquid upon cooling become more and more viscous to finally fall out of equilibrium and solidify. This happens to all liquids cooled fast enough to avoid crystallization \cite{tam33}. Glass-forming liquids exhibit universal features as regards their physical properties \cite{kau48,bra85,sch86,ang00,deb01,das04,bin05,dyr06}. Thus whether the liquid atoms or molecules interact by covalent, ionic, van der Waals, hydrogen, or metal bonds, highly viscous liquids are difficult to distinguish by measurement of macroscopic physical quantities. Glass-forming liquids are characterized by the {\it three non's}: {\it Non}-exponential relaxation functions (linear, as well as nonlinear), {\it non}-Arrhenius temperature dependence of the average relaxation time, and {\it non}linearity of relaxations upon finite, but small perturbations (e.g., a $0.5$ \% temperature change). 

The three {\it non's} are usually regarded as the most basic characteristics of highly viscous liquids (i.e, liquids with viscosity $10^{10}-10^{15}$ times that of ambient water). Two facts are possibly even more fundamental, however: a) Glass-forming liquids are generally supercooled and out-of-equilibrium in the sense that the crystalline phase has lower free energy; b) Glass-forming liquids have extremely large viscosity just above the glass transition (thus justifying the term ``ultraviscous'' used occasionally by Angell, Johari, Stanley, and others).

According to the traditional picture the barrier to crystallization has two contributions \cite{turnbull,deb96}, a barrier for (homogeneous) {\it nucleation} of crystallites of critical size, and a barrier for {\it growth} of these crystallites into macroscopic crystals. The growth barrier is controlled by viscosity and increases upon cooling. The barrier for nucleation, on the other hand, decreases upon cooling. There are several critiques and modifications of this simple picture \cite{deb96,modifications}, but since the below reasoning refers to respectable crystallite sizes, we shall nevertheless adopt its basic logic.

In the classical picture the critical crystallite radius $r_c$ above which growth is automatic and irreversible, is calculated as follows \cite{turnbull}. The crystallite free energy has two contributions, a surface term $4\pi r^2\gamma$ where $r$ is the radius and a volume term $(4\pi/3) r^3\Delta\mu$ where $\Delta\mu<0$ is the free energy difference per unit volume between the crystalline and liquid phases. The latter quantity increases numerically upon cooling, starting from zero at the melting temperature $T_m$. The critical radius is the local maximum of the free energy as function of radius, thus $r_c=2\gamma/|\Delta\mu|$. Note that $r_c\rightarrow\infty$ as $T\rightarrow T_m$.

Most molecular motion in highly viscous liquids consists of vibrations; only rarely do the molecules rearrange by a so-called flow event. Ultraviscous liquids are much like disordered solids, albeit solids that are able to flow over long times if subjected to external forces. The property of being like a solid, an idea that is as old as the research field itself \cite{kau48,gol69}, was termed ``solidity'' \cite{solidity}. We shall argue that the solidity of ultraviscous liquids has important consequences for the occurrence of relatively large subcritical crystallites. The ``solid'' state defined by a potential energy minima -- a so-called inherent state \cite{sti83} -- corresponds to a disordered solid, so there are large frozen-in stresses \cite{ale98}. Because we are dealing with a liquid, these ``frozen-in'' stresses are not permanent, but relax on the alpha time scale. Nevertheless, we shall use this terminology to indicate their very long life.

In ultraviscous liquids the stress tensor is not diagonal, but the pressure $p$ is still well defined via $p\equiv-{\rm tr} (\sigma)/3$. The fluctuation-dissipation theorem implies that at zero external pressure the equilibrium (equal-time) pressure fluctuations determine the instantaneous isothermal bulk modulus according to $K_\infty=(1/k_BT)\int \langle p({\bold 0})p(\br)\rangle d\br$. We shall assume that, as in granular media, the frozen-in stresses are almost spatially random with correlation length $\xi$ of order the intermolecular distance $a$: $\xi\sim a$. The magnitude of the stress fluctuations may then be estimated via $K_\infty\sim\langle p^2\rangle a^3/k_BT$. Typical values of $K_\infty$ are in the range $10^{10}-10^{11}$ Pa s; if we take $a=3$ Angstrom and $T=200$ K, this implies locally fluctuating pressures larger than $10^9$ Pa s. These pressures have two contributions: the kinetic contribution proportional to temperature (that of an ideal gas of same density) and the contribution from molecular interactions. Since $K_\infty$ experimentally always increases upon cooling, the latter contribution must be at least of the same order of magnitude as the kinetic contribution. Thus we estimate that glass-forming liquids have frozen-in pressures of order $10^9$ Pa s. We shall assume that the liquid is at ambient pressure, i.e., basically $p=0$ on the scale set by the frozen-in pressures. From the fact that there is little effective motion it follows that the stress tensor is very close to having zero divergence. It is easy to show that consequently the spatial average pressure is zero, so at any given time there are large positive as well as large negative frozen-in pressures.

In the standard calculation of the crystallite free energy at ambient pressure there is no contribution from the $p\Delta V$ term because $p=0$. Because of the large frozen-in pressures, however, there {\it is} a contribution to the crystallite free energy of $p\Delta V$ origin: Whenever $p>0$ there is a thermodynamic driving force promoting crystallization, but when $p<0$ there is an opposite force tending to melt any crystallite, and these forces do not sum to zero: If $\Delta V<0$ is the difference between crystallite volume and corresponding liquid volume, $\Delta V$ depends significantly on pressure at the enormous frozen-in pressures. At high positive pressure $\Delta V$ becomes numerically smaller than its zero-pressure value, at equally large negative pressure $\Delta V$ increases numerically (more, in fact, because the attractive forces are considerably weaker than the repulsive). Although $\langle p\rangle=0$, there is thus an average contribution $\langle p\Delta V\rangle >0$ to the crystallite free energy. The fact that pressure varies almost randomly in space doesn't affect this argument.

How large is the extra crystallite free energy? A rough analysis is the following: Taking glycerol as an example, the frozen-in pressures ($\sim 10^9$ Pa s) are almost a factor of $10$ larger than the kinetic pressure. Thus if the density difference between liquid and crystal is of order $10\%$, the extra crystallite free energy is of order $k_BT$ per molecule, which is comparable to the typical free energy difference between the crystalline and liquid phases. This means that the $p\Delta V$ contribution cannot be ignored. If the pressure correlation length is temperature independent, the frozen-in pressures increase upon cooling because so does the instantaneous bulk modulus. This increase may be quite pronounced \cite{dyr06}, and it seems possible that in some cases the total bulk free energy of formation upon supercooling develops a maximum as function of temperature. If this happens, there is a temperature where the critical crystallite radius is minimal, whereafter it starts to increase upon continued supercooling. In this case unstable crystallites could be fairly large just above the calorimetric glass transition. (Grinfeld in 1986 \cite{gri86} showed that in the absence of surface tension a plane interphase boundary between a crystal and its melt is unstable with respect to perturbations of any wavelength if the crystal is subjected to non-hydrostatic stresses (see also Ref. \cite{noz93}). This is not the effect discussed here, though.)

The above considerations suggest the following picture: A supercooled liquid consists of a (dominant) liquid phase in which there are well-defined crystallites, some of which are fairly large. If and when the latter by fluctuation get sufficiently large, the liquid crystallizes irreversibly, of course. But if the critical radius is relatively large, there is a high degree of stability of the supercooled liquid phase -- a system in a quasi-static equilibrium where crystallites continuously form and melt away. 

A possible objection against this would be that this situation should have been observed long ago in X-ray and neutron experiments. But that is not necessarily correct. First of all, if only a small fraction of the molecules form crystallites, their signal is weak. Specifically, consider a ``$1/N$-model'' where $1/N$ of the molecules form $N\times N\times N$ crystallites where $N$ could realistically be, e.g., between $5$ and $20$. The crystallites are not only few, but the frozen-in stresses make them even more difficult to observe because these stresses are large enough to deform the crystallites considerably. As a simple example, consider an $N$ crystallite in one dimension with lattice constant $a$. Its contribution to the static structure factor is proportional to $|1+...+\exp(ikNa)|^2$ which is $\sin^2(k(N+1)a/2)/\sin^2(ka/2)$. This gives rise to a signal which is, in principle, observable at arbitrarily large $k$ vectors. If, however, the lattice constant $a$ varies slightly, the average scattering signal gets blurred \cite{mu98,mil00}. Thus if, e.g., the crystallites have a lattice constant that varies a factor $1/N$, there would be no clearly visible contribution to the static structure factor above the liquid signal from the remaining $(N-1)/N$ fraction of the molecules. The same reasoning applies in three dimensions, where shear deformations of course supplement the purely bulk deformations.

Hosemann long ago suggested ``paracrystals'' as a generic model of condensed matter \cite{hosemann}. A paracrystal is a crystal with lattice constants varying randomly over some narrow range. This idea may appear unphysical since lattice constants are not mathematical parameters, but determined by well-defined interaction potentials. The above picture of a glass-forming liquid containing randomly deformed crystallites, however, to some extent realizes Hosemann's idea. The difference is that the crystallites in glass-forming liquids are not permanent -- they are rather ``dynamic paracrystals'' -- and that, moreover, only a minor fraction of the liquid molecules participates in this structuring. 

What are the possible experimental signatures of crystallites? 1) Although virtually impossible to observe by standard diffraction techniques, the higher than liquid average density of crystallites gives rise to a minor excess contribution to small-angle scattering signals and to light scattering. 2) At very high pressures the above picture changes qualitatively, because the absence of negative pressures makes large crystallites unstable. Thus crystallization should be induced by application of very high pressure to a well-annealed ultraviscous liquid if temperature  is simultaneously increased to keep the relaxation time constant. 3) The very low-frequency purely Debye contribution to the dielectric relaxation of monohydroxy alcohols (see, e.g., Ref. \cite{ranko} and its references) could arise from rotation of ferroelectric crystallites. 4) Water is peculiar by having crystal density lower than that of the liquid phase. The above considerations thus do not apply for water, which is predicted to be an anomalous glass former. Indeed, glassy water is known to be unstable against crystallization. 5) Ultraviscous liquids without a crystalline phase of lower free energy should behave differently from ordinary glass-forming liquids.

The idea that supercooled liquids contain crystallites is as old as glass science itself. Many years ago it was suggested that glass consists of very small crystallites (obviously, these must be assumed to be present in the melt). This was later referred to as the ``old crystallite hypothesis'' \cite{wri98} (see also Ref. \cite{hen05}). More refined  is the ``new crystallite hypothesis'' of Valenkov and Porai-Koshits who in a 1936-paper on glass structure \cite{val36} wrote: ``A hypothesis is made, according to which the crystallites are consisting of an inner part with regularly deformed (stretched or compressed) space lattice, and of outer portions with strong and arbitrary distortions.'' Again the glass is assumed to consist entirely of crystallites, but now the crystallites are thought of as gradually merging with neighboring crystallites by being deformed in their outer portions. Wright noted \cite{wri98} that ``the problem is to explain why such nuclei/crystallites should be so stable, but not grow'' \cite{note1}. The $p\Delta V$ contribution to the crystallite free energy provides a possible explanation for why growth is inhibited, because this term lowers the bulk free energy gain by crystallite formation and thus increases the critical crystallite radius. -- Recently, Tanaka in an interesting series of papers \cite{tanaka} proposed a ``two-order-parameter'' model according to which the liquid-glass transition is controlled by competition between long-range crystalline ordering and short-range ordering into locally favored structures incompatible with crystalline order. In this model there are ``metastable solid-like islands'' in the liquid with a ``hidden crystalline order'' and there are ``locally favored structures;'' the alpha relaxation is associated with the creation and annihilation of the metastable islands. The present picture may be regarded as the logical extrapolation of Tanaka's by having a liquid phase ($\sim$ his ``locally favored structures'') and a crystalline phase ($\sim$ his ``solid-like islands''). In terms of order parameters \cite{op} in the present picture there is just one natural order parameter, namely the fraction of molecules that form crystallites \cite{note2}.

To summarize, we discussed the possibility that ultraviscous glass-forming liquids may contain few, but fairly large and well-defined crystallites that continuously form and melt away. We are aware that the above reasoning is highly speculative \cite{note3}. Thus if for instance the pressure fluctuations have larger correlation length than assumed, the frozen-in pressures are not large enough to contribute significantly to the crystallite free energy. Here we merely wished to draw attention to an interesting possibility which -- if valid -- should be important for understanding the physical properties of ultraviscous liquids.

\section{Acknowledgments}
This work was supported by the Danish National Research Foundation Centre for Viscous Liquid Dynamics ``Glass and Time.''


\begin{thebibliography}{99}
\bibitem{note3} There are no plans to submit this note to a regular journal (in its present form, at least).
\end{thebibliography}
\end{document}